\documentclass[a4paper,10pt]{article}
\usepackage[a4paper, total={5.5in, 8in}]{geometry}
\usepackage[utf8]{inputenc}
\usepackage{graphicx}

\def\nn {\nonumber}

\newcommand{\be}{\begin{equation}}
\newcommand{\ee}{\end{equation}}
\newcommand{\bea}{\begin{eqnarray}}
\newcommand{\eea}{\end{eqnarray}}

\newcommand{\om}{\omega}
\newcommand{\vk}{\vec k}
\newcommand{\vq}{\vec q} 

\newcommand{\mn}{\mu\nu}
\newcommand{\del}{\partial}

\newcommand{\lnz}{\ln\mathcal{Z}}

\newcommand{\munu}{{\mu\nu}}

\newcommand{\FB}[1]{\left(#1\right)}

\begin{document}



\title{Viscosity calculations from Hadron Resonance Gas model: Finite size effect}
\date{}
\author{Snigdha Ghosh$^1$, Subhasis Samanta$^2$, Sabyasachi Ghosh$^3$, 
\\
and Hiranmaya Mishra$^4$}
\maketitle
\begin{center}
{$^1$Indian Institute of Technology Gandhinagar, Palaj, 
Gandhi nagar 382355, Gujarat, India}
\\
{$^2$School of Physical Sciences, National Institute of Science Educa
tion and Research, Bhubaneswar, HBNI, Jatni, 752050, India}
\\
{$^3$Indian Institute of Technology Bhilai, GEC Campus, Sejbahar, 
Raipur-492015, Chhattisgarh, India}
\\
{$^4$Theory Division, Physical Research Laboratory, Navrangpura, Ahmedabad 380 009, India}
\end{center}
%
%

\begin{abstract}
We have attempted to review on microscopic calculation of transport coefficients
like shear and bulk viscosities in the framework  of hadron resonance gas model,
where a special attention is explored on the effect of finite system size.
The standard expressions of transport coefficients, obtained from relaxation time
approximation of kinetic theory or diagrammatic Kubo-type formalism, carry mainly two
temperature dependent components - thermodynamical phase space and relaxation time of medium constituent.
Owing to quantum effect of finite system size, thermodynamical phase space can be
reduced as its momentum distribution will be started from some finite lower momentum
cut-off instead of zero momentum. On the other hand, relaxation time of hadrons can also
face finite size effect by considering only those relaxation scales, 
which are lower than the system size. Owing to these phenomenological issues, we have proposed 
a system size dependent upper bound of transport coefficients for ideal HRG model, 
whose qualitative technique may also be applicable in other models. This finite size prescription may guide
to shorten the broad numerical band, within which earlier estimated values of transport 
coefficients for hadronic matter are located. It is also suspected that the hadronic matter
may not be far from the (nearly) perfect fluid nature like the quark gluon plasma. 

%
\end{abstract}




\section{Introduction}

The ratio between shear viscosity $\left(\eta\right)$ and entropy density $(s)$ of any medium 
is the quantity, which measures the fluidity of the medium. 
This ratio $\eta/s$ is roughly equal to the ratio of mean free path to de-Broglie wavelength of 
medium constituent. Hence, it can never be vanished, as mean free path 
of any constituent can never be lower than its de-Broglie wavelength. Other way to comment
is that quantum fluctuations prevent the existence of perfect fluid in nature and $\eta/s$ of 
any fluid should have some lower bound, which is also claimed from the 
string theory calculation~\cite{KSS}. 

%

Interestingly, a small value of $\eta/s$, very close to this quantum lower bound, is observed
in super hot medium, produced in heavy ion collision experiment like RHIC~\cite{RHIC_kss} and LHC~\cite{LHC_kss}.
Some other many body systems like cold atoms~\cite{coldatom}, graphene~\cite{graphene} and in
low energy nuclear matter~\cite{VECC} are also observed to have small $\eta/s$ near to the lower bound. 
This nearly perfect fluid behavior at different extreme situations has drawn 
immense attention from scientific communities working on the field of 
condense matter physics to nuclear physics to string theory.

Present article is interested on fluid properties of hadronic matter, where a special
phenomenological attention is taken care on finite size of the system.
One can find a long list of Refs.~\cite{Prakash,Itakura,Dobado,Nicola,Weise1,SSS,Ghosh_pi,Ghosh_piN,SS_piN,
Gorenstein,Denicol,HM1,HM2,Hostler,SG_HRGV_JPG,SG_HRG_PRC,Bass,Muronga,Pal,Juan} on microscopic 
calculations of $\eta/s$ for hadronic matter,
where some of them adopted different 
hadronic models~\cite{Prakash,Itakura,Dobado,Nicola,Weise1,SSS,Ghosh_pi,Ghosh_piN,SS_piN,
Gorenstein,Denicol,HM1,HM2,Hostler,SG_HRGV_JPG,SG_HRG_PRC}
and some have gone through bulk simulations~\cite{Bass,Muronga,Pal,Juan}. 
Their predicted values of $\eta/s$ reside within a broad numerical band. 
Similar kind of broad numerical band is also observed for bulk viscosity 
$\zeta$ of hadronic matter, estimated by earlier Refs.~\cite{{Nicola},{HM1},{HM2},{Hostler},{Purnendu},{Prakash},{Gavin},
{Dobado_zeta1},{Dobado_zeta2},{Nicola_PRL},{Sarkar},{SG_NISER},{Sarwar}}.
Recently, Refs.~\cite{SG_HRG_PRC,SG_HRGV_JPG} have provided a possible prescription
to make the band be narrow, when a finite size effect of medium is considered.
Though the prescription is quantitatively applicable in hadron resonance gas (HRG) model,
adopted in Refs.~\cite{SG_HRG_PRC,SG_HRGV_JPG}, but a qualitative application of this prescription
can always be possible in other hadronic models.
Present article is intended to review this prescription by analyzing earlier estimated
values of $\eta/s$ and $\zeta/s$. 

Starting with a brief formalism of transport 
coefficients (\ref{Sec:transport coefficients}) and HRG model (\ref{Sec:HRG}) parts in 
Sec.~(\ref{Sec:Formalism}), we have explored the results of finite system size effect
in next Sec.~(\ref{Sec:LMC}), which contain the discussion
about a lower momentum cut-off consideration in thermodynamical quantities and its 
phenomenological aspect~\cite{SG_HRGV_JPG}. Next Sec.~(\ref{Sec:Relaxation}) describe the relaxation
time of different hadrons, which are classified in two components, discussed in two subsections
(\ref{Sec:NR}) and (\ref{Sec:R}). An upper bound is proposed for finite size of system~\cite{SG_HRG_PRC}.
Considering a special example in Sec.~(\ref{Sec:Example}), we show how an upper
momentum cut-off in relaxation scale help to reduce the values of $\eta/s$ and $\zeta/s$
below their upper bound~\cite{SG_HRG_PRC}. At the end in Sec.~(\ref{Sec:Summary}), we have concluded. 
%

%


\section{Formalism}
\label{Sec:Formalism}
\subsection{transport coefficients}
\label{Sec:transport coefficients}
%
%
The proportional constants between thermodynamical forces and fluxes are basically defined as transport coefficients. 
Here, we will discuss about the two transport coefficients - shear and bulk viscosities, although other transport 
coefficients like electrical and thermal conductivities may also be very interesting to study.
We start with the energy-momentum tensor of a relativistic fluid $T^\munu$ which is split into a 
ideal and an dissipative part as:
\begin{eqnarray}
T^\munu = T_0^\munu + T_D^\munu~,
\end{eqnarray} 
where, $T_0^\munu = (\epsilon+P)u^\mu u^\nu - Pg^\munu$, containing energy density $\epsilon$,
pressure $P$, fluid four-velocity $u^\mu$, metric tensor $g^{\mu\nu}$.
Ignoring the heat flow part, the dissipation part of energy momentum tensor will be
\begin{eqnarray}
T_D^\munu = \eta U^{\mn}_\eta + \zeta U^{\mn}_\zeta
\label{eq.Pimunu}
\end{eqnarray}
where thermodynamical fluxes for shear ($\eta$) and bulk ($\zeta$) viscosities are 
\bea
U_\eta^{\mn} &=& D^\mu u^\nu + D^\nu u^\mu +\frac{2}{3}
\Delta^{\mu\nu}\partial_\sigma u^\sigma
\nn\\
U_\zeta^{\mn} &=& \Delta^\munu \FB{\del^\alpha u_\alpha}
\eea
with $D^\mu=\partial^\mu - u^\mu u^\sigma \partial_\sigma$ and $\Delta^\munu = g^\munu - u^\mu u^\nu$. 

Here, we are going to discuss about two different approaches for the evaluation of $\eta$ and $\zeta$ namely: 
(i) kinetic theory approach and (ii) Kubo approach. 

In former approach, we assume that each hadrons in the medium is
deviated from their equilibrium distribution function $f^0_h= 1/\left[ e^{\om_h/T} +a_h \right]$ by a 
small amount $\delta f=\phi f^0_h(1+a_h f^0_h)$, where
$a_h = \pm 1$ for boson and fermion respectively;
$\phi=a_\eta k_\mu k_\nu  U^{\mn}_\eta + a_\zeta  U^{\mn}_\zeta $ is assumed in terms of same tensor 
decomposition as given in Eq.~(\ref{eq.Pimunu}).
In terms of these quantities, the dissipative part of energy-momentum tensor,
\be
T^{\mn}_D=\sum_{h} g_h\int \frac{d^3\vk}{(2\pi)^3}\frac{k^\mu k^\nu}{\om_h}\delta f~,
\ee
can be obtained in the form of Eq.~(\ref{eq.Pimunu}), where the expressions of $\eta$ and $\zeta$ will come
in term of unknown coefficients $a_\eta$ and $a_\zeta$. Now, using the relaxation time approximation of
relativistic Boltzmann transport equation,
\begin{eqnarray}
p_h^\mu\del_\mu f_h(x,p_h)=-p_h^0\frac{\delta f_h}{\tau_h}~,
\label{eq.BTE}
\end{eqnarray}
the unknown coefficients $a_\eta$, $a_\zeta$ can be obtained and we will get RTA expression of $\eta$ and $\zeta$ 
as~\cite{Purnendu,Gavin}
\begin{eqnarray}
\eta&=&\sum_{h}\frac{g_h}{15 T}\int \frac{d^3\vk}{(2\pi)^3}
\tau_{h}\left(\frac{\vk^2}{\om_{h}}\right)^2\frac{}{}f^0_h\left(1-a_hf^0_h\right)~, 
\label{eta_G}
\\
\zeta&=&\sum_{h}\frac{g_h}{T}\int \frac{d^3\vk}{(2\pi)^3\om_{h}^2}
\tau_{h}\left\{\left(\frac{1}{3}-c_s^2\right)\vk^2  - c_s^2 m_h^2\right\}^2
f^0_h\left(1-a_hf^0_h\right)~,
\label{zeta_G}
\end{eqnarray}
where $c_s$ is the velocity of sound in the medium and $\tau_h$ is relaxation time of $h$ hadron.
The $g_h$ is degeneracy factors of hadrons due to spin, isospin (including particle, anti-particle
counting as we are interested on zero baryon density picture). 

Similar kind of expressions can be obtained in Kubo approach, where transport coefficients are defined
as zero energy-momentum limit of correlators, constructed by energy momentum tensor. 
The standard (field theoretical version) Kubo expressions of $\eta$ and $\zeta$ 
are~\cite{Nicola,Weise1,G_IJMPA}
\bea
\eta&=&\frac{1}{20}\lim_{q_0,\vq\rightarrow 0}
\frac{\int d^4xe^{iq\cdot x}\langle [\pi^{ij}(x),\pi_{ij}(0)]\rangle_\beta}{q_0}~,
\nn\\
\zeta&=&\frac{1}{2}\lim_{q_0,\vq\rightarrow 0}
\frac{\int d^4xe^{iq\cdot x}\langle [{\cal P}(x),{\cal P}(0)]\rangle_\beta}{q_0}~,
\eea
where the operators $\pi_{ij}$ and ${\cal P}$ can be obtained from energy-momentum tensor $T^{\mn}$
as,
\bea
\pi^{ij}&\equiv&T^{ij}-g^{ij}T^k_k/3~,
\nn\\
{\cal P}&\equiv&-T^k_k/3 -c_s^2T^{00}~.
\eea 
When one builds $T^{\mn}$ from free Lagrangian density, the correlators take the form in terms of
boson or fermion fields and then we will get a simplest one-loop skeleton level diagram, which is
divergent in zero energy-momentum limit. To cure this divergence, a finite thermal width ($\Gamma_h=1/\tau_h$) is
generally introduced in the propagators of one-loop diagram and at the end of the 
calculation~\cite{Nicola,Weise1,G_IJMPA}, we will get 
same expressions of $\eta$ and $\zeta$, as obtained in RTA method, given in Eqs~(\ref{eta_G}), (\ref{zeta_G}).

\subsection{Thermodynamical quantities from HRG model}
\label{Sec:HRG}
In this work, we aim to calculate the quantities $\eta/s$ and $\zeta/s$ from 
Eqs.~\ref{eta_G} and \ref{zeta_G} for which the necessary inputs are the 
thermodynamic quantities: entropy density $s$ and speed of sound $c_s$ of the medium.
A unified thermodynamics of quark and hadronic matter with a smooth cross-over quark-hadron
transition can be obtained from lattice QCD simulations~\cite{Borsanyi:2013bia,LQCD_2014}.
%
It is remarkable that, QCD thermodynamics in the low temperature region ($100 < T < 160$) MeV 
can achieved analytically by considering a grand canonical ensemble of all the 
non-interacting hadrons- a formalism known as Hadron Resonance Gas (HRG) Model~\cite{HRGrev}.
We start with the grand canonical partition function at zero chemical potential ($\mu=0$)
%
\be
\ln\mathcal{Z}\left(T,V,\mu=0\right) = V\int\frac{d^3p}{(2\pi)^3}\sum_{h}g_ha_h 
\times \ln\left[1+a_h
\exp\left\{-\beta\omega_h\right\}\right]~.
\label{eq.partition.function}
\ee
%
%
All the relevant thermodynamic quantities like pressure ($P$), energy density ($\varepsilon$), 
entropy density ($s$) and speed of sound ($c_s$)
are obtained from the partition function as:
\bea
s &=& \left(\frac{\varepsilon+P}{T}\right)=\frac{1}{T}\left\{\left(\frac{T}{V}\right)\lnz
+\left(\frac{T^2}{V}\right)\frac{\del}{\del T}\left(\lnz\right)\right\}
\label{eq.entropy}
\\
 c_s^2 &=& \left(\frac{\partial p}{\partial\epsilon}\right) 
 = \left(\frac{\partial p}{\partial T}\right)\Big/\left(\frac{\partial\epsilon}{\partial T}\right) 
 \label{eq.cs2}~.
\eea
Now, using the Eq.~(\ref{eq.partition.function}) in Eqs.~(\ref{eq.entropy}) and (\ref{eq.cs2}),
one can obtain $s$ and $c_s^2$, which is presented by red solid line in Fig.~\ref{fig:s_sSB_T}(a)
and (b) respectively. The entropy density has been normalized by its Stefan-Boltzmann (SB) limit 
$s_{SB}=\frac{19\pi^2}{9}T^3$ (for 3 flavor quarks).
The LQCD data of $s/s_{SB}$ and $c_s^2$ from the WB group~\cite{Borsanyi:2013bia} (triangles)
and Hot QCD group~\cite{LQCD_2014} (circles) are also plotted in Fig.~\ref{fig:s_sSB_T}(a)
and (b), which are in well agreement with the red solid curves, obtained
in ideal HRG model. This friendship between HRG model and LQCD within the hadronic temperature
range is well familiar fact.
\section{Finite size: Lower momentum cut-off}
\label{Sec:LMC}
\begin{figure}
\includegraphics[width=0.48 \textwidth]{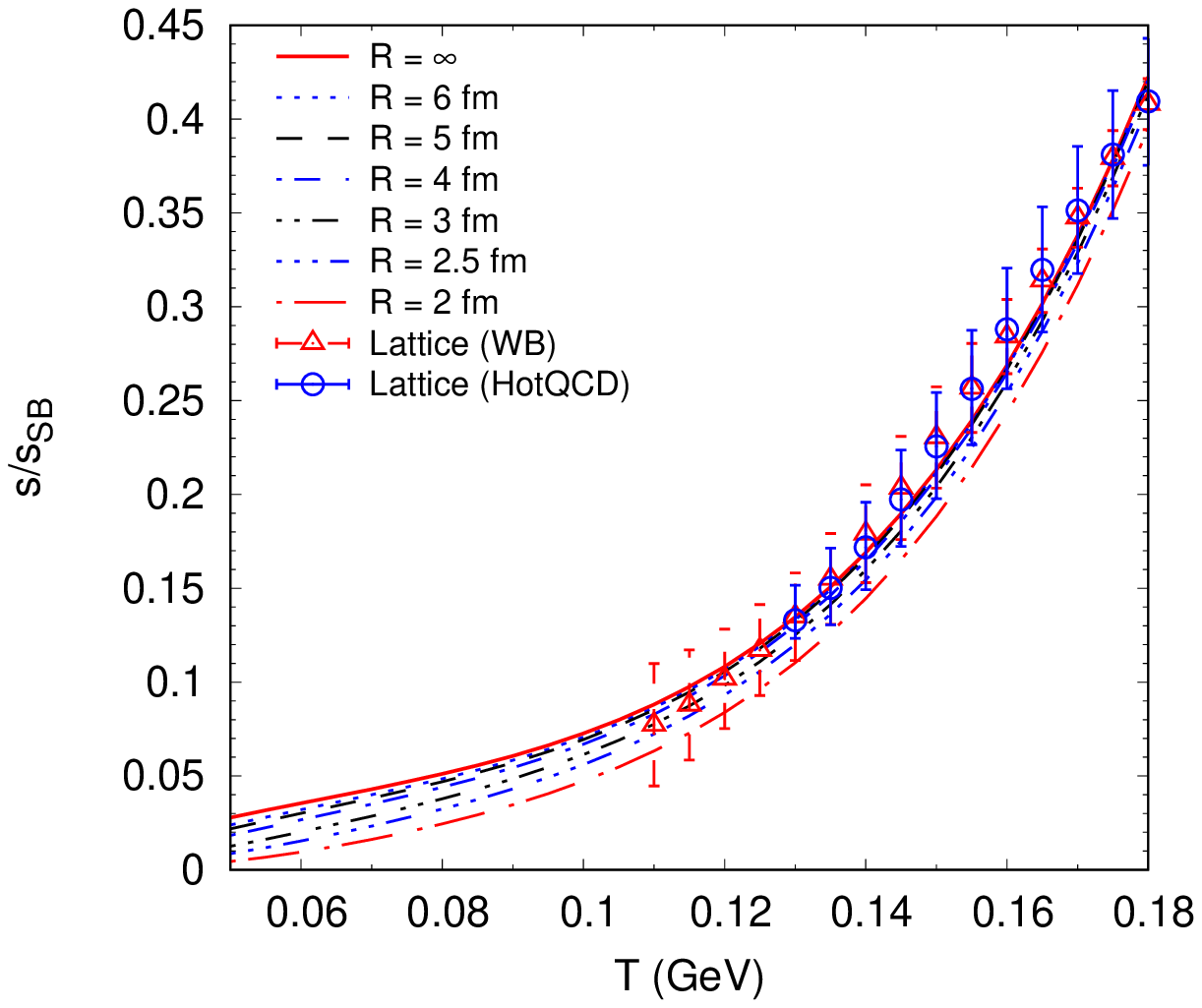}
\includegraphics[width=0.48 \textwidth]{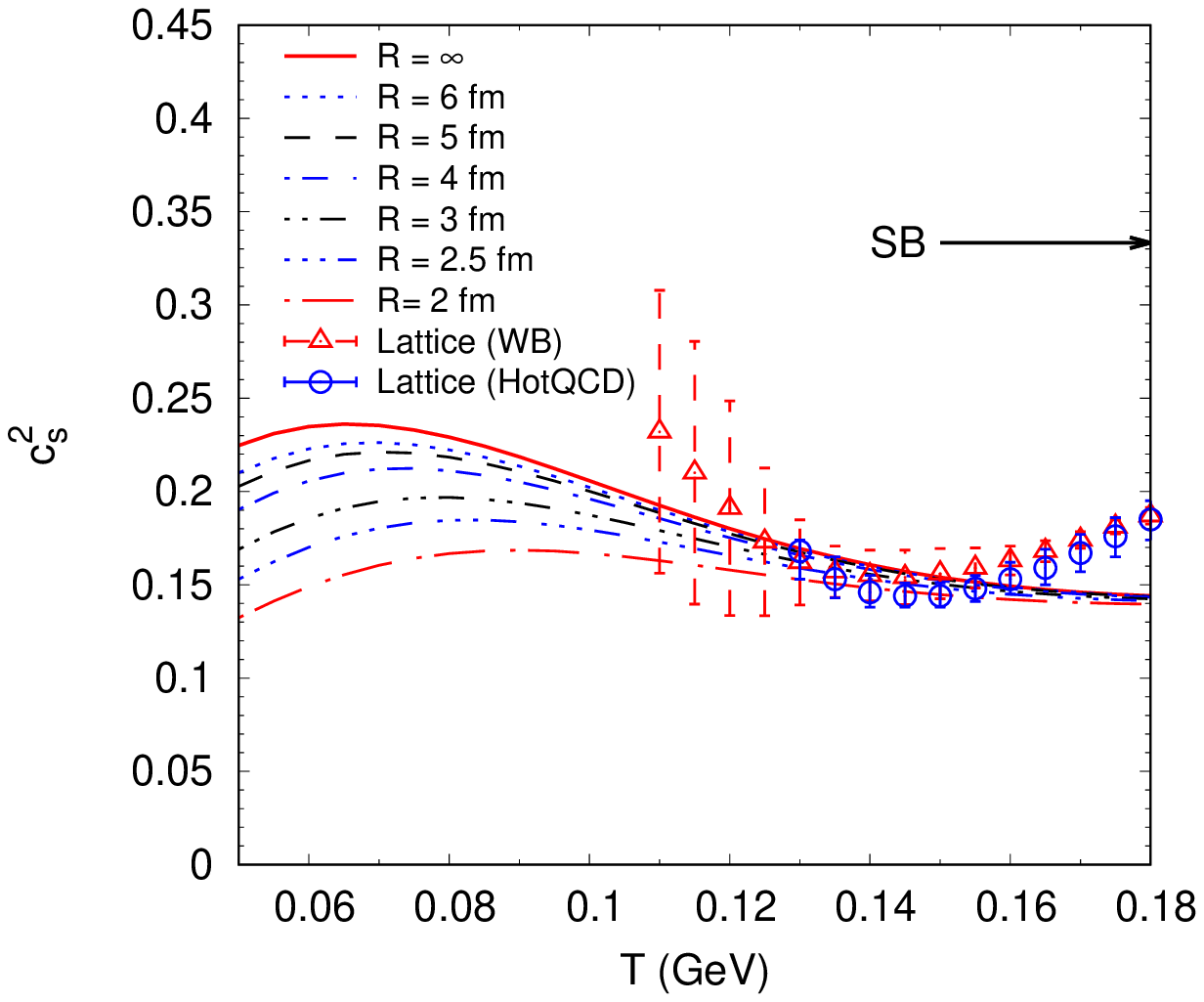}
\caption{(Color online) (a) The ratio of entropy density to its Stefan-Boltzmann (SB) 
value ($s/s_{SB}$) is plotted against $T$ for different values of $R$. 
(b) Square of speed of sound ($c_s^2$) versus $T$ for different values of $R$.
The Stefan-Boltzmann limit of $c_s^2$ is shown by the arrow. Corresponding 
LQCD data from the WB group~\cite{Borsanyi:2013bia} (triangles) and Hot QCD group~\cite{LQCD_2014} 
(circles) are also shown.}
\label{fig:s_sSB_T}
\end{figure}
Here we want to shift our focus on some phenomenological point,
where we will consider that the hadronic medium, produced in heavy ion collision experiments,
has some finite size. So idea is to modify the theoretical tools of (ideal) HRG model for
applying in realistic scenario, where the hadronic matter carry a finite size.

Reader can find a vast literature
\cite{Luscher:1985dn,Gasser:1987ah,Spieles:1997ab,
Gopie:1998qn,Abreu:2006pt,Luecker:2009bs,Fraga:2011hi,Bhattacharyya:2015zka, Bhattacharyya:2015kda,Magdy:2015eda,
Redlich:2016vvb,RedlichV,KinkarV}
on finite size method, where we will consider the simplest toy model, as adopted in
Refs.~\cite{Bhattacharyya:2015zka,Bhattacharyya:2015kda,KinkarV}~.
If the volume is so small that the quantum effect can not be ignored then 
one can roughly relate the system with the quantum mechanical picture for particle in a box. 
Hence, the lowest possible zero momentum in classical picture will be transformed to
a finite value in quantum picture, which depends on the size 
as~\cite{Bhattacharyya:2015zka,Bhattacharyya:2015kda,KinkarV}
\be
k_{\rm min}=\pi/R~,
\label{kmin_R}
\ee
where $R$ is the size of a cubic volume $V\sim R^3$ of
the system. Let us call it lower momentum cut-off. 
So, implementing the transformation
\be
\int^{\infty}_0 \rightarrow \int^{\infty}_{k_{\rm min}}
\label{0_Kmin}
\ee
in Eq.~(\ref{eq.partition.function}), one can get a partition function of finite size hadronic matter.
In principle, the integration should also be converted to sum over discrete momentum values but for
simplicity we have not considered that.
This simplified approximated picture is well justified in Refs.~\cite{Redlich:2016vvb,RedlichV}.
%
We have taken six different sizes of $R = 2, 2.5, 3, 4, 5, 6$ fm, whose
range is guided from the experimental values, provided in Ref.~\cite{STAR_17}~.
In Figs.~\ref{fig:s_sSB_T}(a-b), we notice  that $s$ and $c_s^2$ both are decreasing as $R$ decreases. It 
indicates that the values of $s$ and $c_s^2$ for theoretically assumed infinite
hadronic matter can not be considered for experimentally produced medium, having
a finite size.

%

\subsection{Phenomenological aspect of lower momentum cut-off}
\label{Sec:pheno}
\begin{figure}
\begin{center}
\includegraphics[width=0.9 \textwidth]{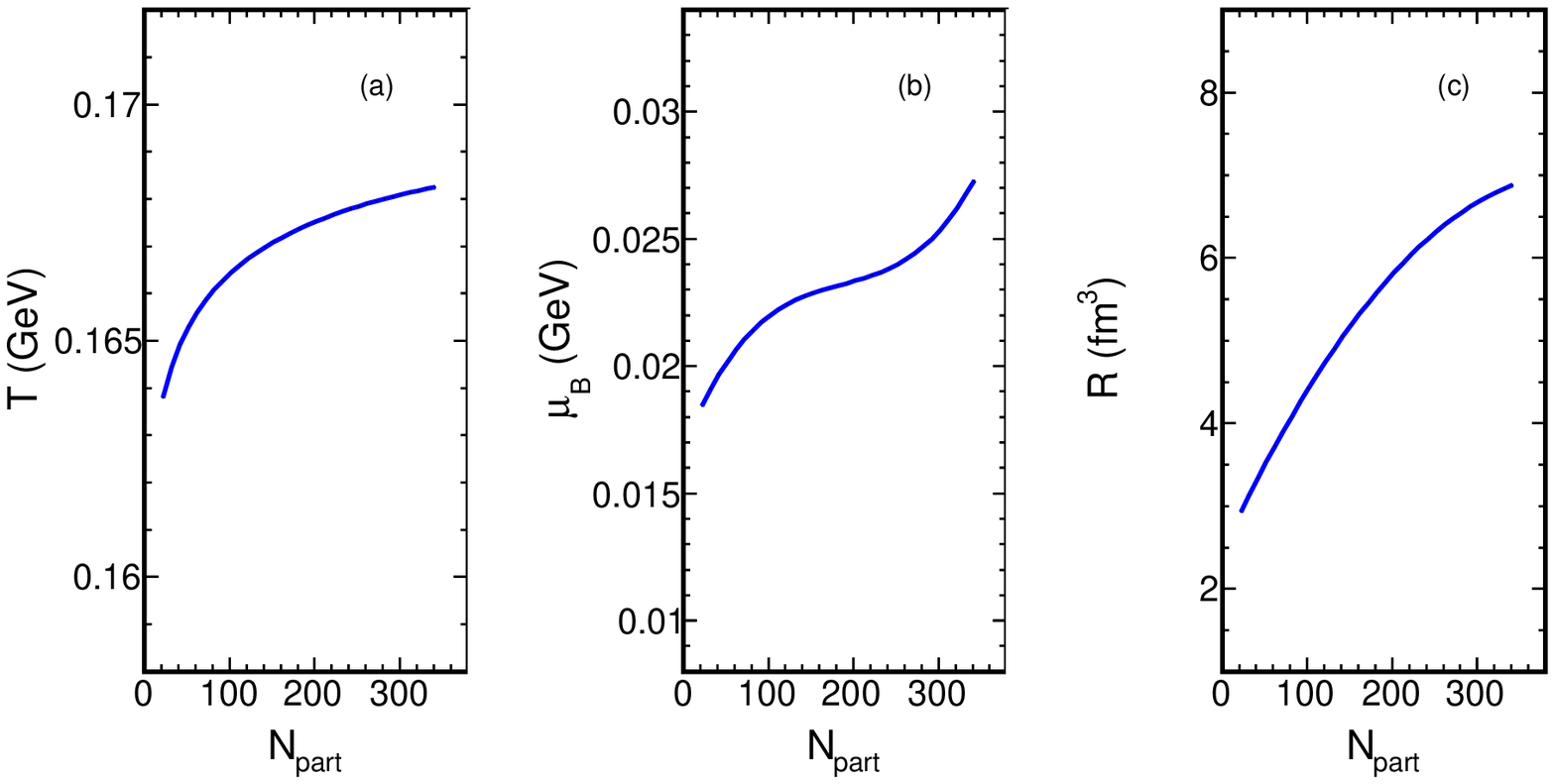}
\includegraphics[width=0.7 \textwidth]{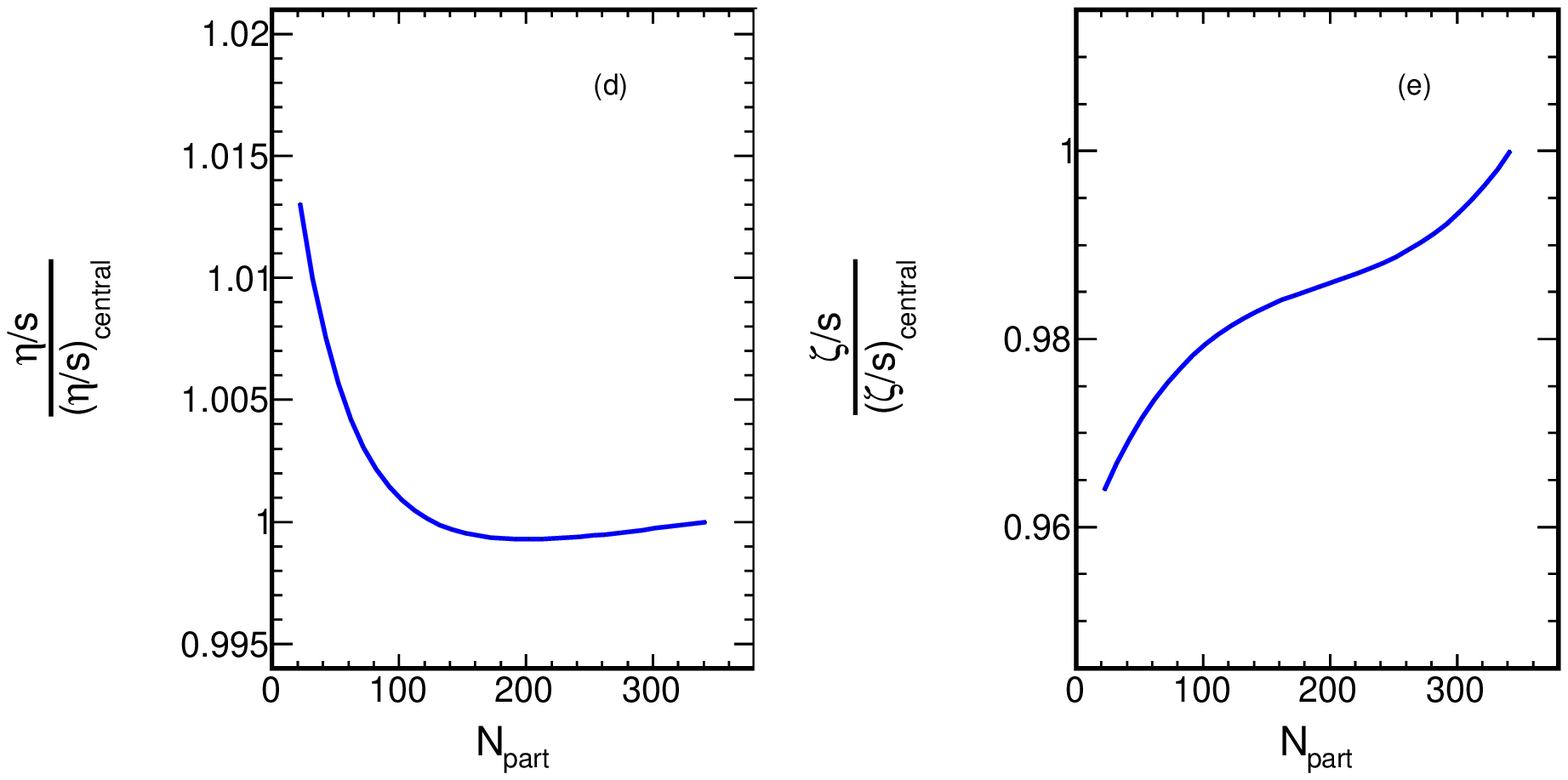}
\caption{$N_{\rm part}$ dependence of temperature $T$ (a), baryon chemical 
potential $\mu_B$ (b) and radius $R$ (c) of the medium at freeze-out point, 
taken from Ref.~\cite{STAR_17}. Using those inputs, the dimensionless transport 
coefficients $\eta/s$ (d), $\zeta/s$ (e) are plotted 
against $N_{\rm part}$.} 
\label{R_npart}
\end{center}
\end{figure}

We have noticed that thermodynamical quantities like $s$, $c_s^2$ are changed due to
lower momentum cut-off $(k_{\rm min})$ consideration, mapping the finite size effect. Similarly,
shear and bulk viscosities will also be changed, when
one transforms the lower limit of integration in Eqs.~(\ref{eta_G}) and (\ref{zeta_G})
from $0$ to $k_{\rm min}$. The detail result have been addressed in Ref.~\cite{SG_HRGV_JPG}.
Interestingly, we get a link between this microscopic estimation of viscosities for finite 
size system and the macroscopic evolution picture.
%
%
For non-central
collision, the freeze-out size $R$ becomes smaller.
So one can expect different values
of transport coefficients for different centrality,
which is mapped by the number of participant $N_{\rm part}$.
From the table 8 of Ref~\cite{STAR_17} for Au+Au collision at RHIC energy  $\sqrt{S_{NN}} = 200$ GeV, 
we have taken the experimental data of centrality or $N_{\rm part}$ dependence of chemical freeze-out
parameters such as temperature $T$, chemical potential $\mu_B$ and system size $R$.  After
parameterizing the data, the functions $T(N_{\rm part})$, $\mu_B(N_{\rm part})$ and $R(N_{\rm part})$ 
are plotted in Figs.~\ref{R_npart}(a)-(c), which shows roughly the variation $R=7-3$ fm, 
$\mu_B=0.027-0.019$ GeV and $T=0.168-0.164$ GeV for $N_{\rm part}=350-40$. 
So $R$ faces major changes and $T$, $\mu_B$ face mild changes due to change of $N_{\rm part}$.
Using those input values of $T$, $\mu_B$ and $R$ in Eqs.~(\ref{eta_G}) (\ref{zeta_G}) and (\ref{eq.entropy}), 
we have obtained corresponding $N_{\rm part}$ dependence of $\eta/s$
and $\zeta/s$, where $R$ enters in the lower limit of integration by following Eqs.~(\ref{kmin_R}), 
(\ref{0_Kmin}). These $\eta/s$ and $\zeta/s$ are plotted against $N_{\rm part}$ in Figs.~\ref{R_npart}(d) and (e),
where their values have been normalized by the corresponding values
at most central collision, carrying maximum $N_{\rm part}$.
%
%
We notice that $\eta/s$ decreases with $N_{\rm part}$.  Interestingly, similar
trend is expected from hydrodynamical simulation~\cite{Roy:2012jb}, where
$\eta/s$ is entered as an input parameter.
This qualitative agreement between microscopic~\cite{SG_HRGV_JPG} and macroscopic~\cite{Roy:2012jb}
investigations is pointing out the importance of size-dependent transport coefficient calculations.
%
%

\section{Relaxation time}
\label{Sec:Relaxation}
So far we have concentrated on thermodynamical phase space parts of
$\eta$ and $\zeta$ expressions, given in Eqs.~(\ref{eta_G}), (\ref{zeta_G}).
The known values of different hadron masses~\cite{PDG} fix its numerical
strength. Beside this thermodynamical phase space part, another part is
relaxation time $\tau_h$, which will be analyzed in this section.

Based on the phenomenological picture of dissipation, hadrons
may be classified into two categories - non-resonance (${\cal NR}$) and
resonance (${\cal R}$) components. 

\subsection{Non-Resonance (${\cal NR}$) component} 
\label{Sec:NR}
Let us first discuss about ${\cal NR}$ component, then we will come to 
the discussion of ${\cal R}$ component in next subsection.

Long lived particles like 
pseudo-scalar meson nonet and baryon octet may be considered as ${\cal NR}$ members
as they can't decay within the life time of the fireball, produced in HIC experiments. 
For simplicity, we considering only pion, kaon and nucleon as ${\cal NR}$ members as
they are most abundant constituents in the medium.
They participate in dissipation via their strong interaction scattering process.
Due to their dominant contribution in dissipation, many of the earlier works
have focus on pion gas~\cite{Dobado,Nicola,Weise1,SSS,Ghosh_pi} and 
(pion + nucleon)~\cite{Ghosh_piN,SS_piN,Itakura} gas instead of going to all hadrons,
as considered in HRG model~\cite{Gorenstein,Hostler,Denicol,HM1,HM2}.
\begin{figure}
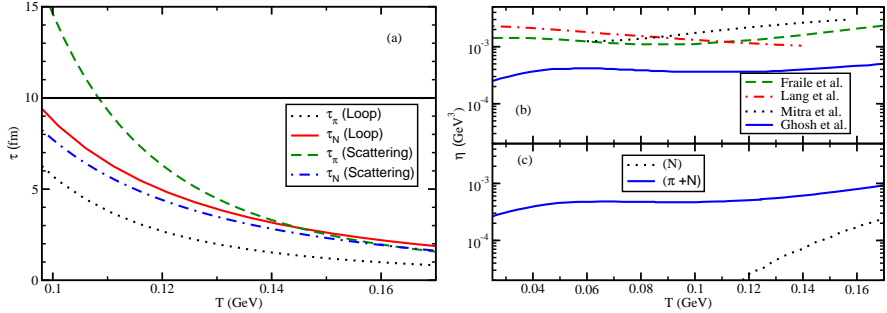
  
\begin{center}
\includegraphics[scale=0.24]{tav_T.eps}
\includegraphics[scale=0.24]{eta_piN_T.eps}
\caption{(Color online) (a) Dotted and dash lines respectively denote the pion relaxation 
time $\tau_\pi$ from loop~\cite{Ghosh_pi} and scattering~\cite{SSS} diagrams calculations; Same 
for nucleon relaxation time $\tau_N$ are displayed by solid~\cite{Ghosh_piN} 
and dash-dotted~\cite{SS_piN} lines respectively. 
(b) Shear viscosities of pion, obtained by Fraile et al.~\cite{Nicola},
Lang et al.~\cite{Weise1}, Mitra et al.~\cite{SSS}, Ghosh et al.~\cite{Ghosh_pi}.
(c) Contribution of nucleon (dotted line) and (pion + nucleon) (solid line) from Ref.~\cite{Ghosh_piN}.}
\label{fig:t_eta_piN} 
\end{center}
\end{figure}
Relaxation time ($\tau_\pi$) or length ($\lambda_\pi$) of pion, obtained from
$\pi^4$-type interaction Lagrangian density, has been considered by Lang et al.~\cite{Weise1}.
Based on the optical theorem of thermal field theory, they have derived the thermal 
width of pion $\Gamma_\pi=1/\tau_\pi$ from the imaginary part of pion self-energy diagram,
containing two-loop with 3 pion propagators. Using the temperature and momentum dependent
relaxation time of pion in Eq.~\ref{eta_G}, they have estimated temperature dependent
shear viscosity of pion ($\eta_\pi$), which is shown by red dash-dotted line 
in Fig.~\ref{fig:t_eta_piN}(b). This leading order interaction Lagrangian density from Chiral
Perturbation Theory (ChPT) can only describe the pion interaction but can't generate
the resonances $\rho$ and $\sigma$ mesons, as experimentally observed in $\pi\pi$ scattering
cross section. These resonances $\rho$ and $\sigma$ mesons can be introduced via phenomenological
$\sigma\pi\pi$ and $\rho\pi\pi$ interaction Lagrangian densities, as considered in Refs.~\cite{Ghosh_pi,SSS}.
Similar resonance picture can also be dynamically generated via unitarization technique, as adopted in Ref.~\cite{Nicola}.
Shear viscosities of pion in resonance picture of Refs.~\cite{Nicola,SSS,Ghosh_pi} have been shown by
green dash, black dotted and solid blue lines respectively in Fig.~\ref{fig:t_eta_piN}. 
Comparing these curves with red dash-dotted lines, we notice that decreasing trend of $\eta_\pi(T)$ in ChPT
can be transformed to increasing nature when one introduces the resonances. This is also checked in the Ref.~\cite{Nicola}. 
Now let us come to other ${\cal NR}$ member - nucleon. Shear viscosities of nucleon $\eta_N$ has been calculated in
Ref.~\cite{Ghosh_piN} via phenomenological $N\pi B$ interaction Lagrangian densities, where $B$ stand for different
spin $1/2$ and $3/2$ baryons. The black dotted line and blue solid lines in Fig.~\ref{fig:t_eta_piN}(c) are respectively
displaying the results of $\eta_N$ and $(\eta_N+\eta_\pi)$ of Ref.~\cite{Ghosh_piN}.
Refs.~\cite{Ghosh_pi,Ghosh_piN} have obtained the relaxation times $\tau_\pi$ and $\tau_N$ from 
the pion and nucleon self-energies, where $\pi\sigma$, $\pi\rho$~\cite{Ghosh_pi} and different $NB$ loops 
are considered for former case, while latter one carries different $\pi B$ loops~\cite{Ghosh_piN},
displayed by black dotted and solid red lines in Fig.~\ref{fig:t_eta_piN}(a). 
Same quantities $\tau_\pi$ and $\tau_N$ in Ref.~\cite{SSS,SS_piN}, shown by dash and dash-dotted 
lines in Fig.~\ref{fig:t_eta_piN}(a), have been obtained from elastic scattering 
processes $\pi\pi\rightarrow\pi\pi$, $\pi N\rightarrow\pi N$ and $NN\rightarrow NN$, 
whose relaxation time scales are $\tau_{\pi\pi}$, $\tau_{\pi N}$ and $\tau_{NN}$. Their mixing relations
are $\frac{1}{\tau_\pi}=\frac{1}{\tau_{\pi\pi}}+\frac{1}{\tau_{\pi N}}$, 
$\frac{1}{\tau_N}=\frac{1}{\tau_{NN}}+\frac{1}{\tau_{N\pi}}$.
The straight horizontal line at 10 fm is assumed
as maximum possible dimension of hadronic matter and we see that the values 
of $\tau_\pi$ and $\tau_N$, obtained from both loop and scattering diagrams, 
are lower than the system size as expected in dissipative fluid picture. 
%
%
%
\subsection{Resonance ($\cal R$) component and total (${\cal NR}+{\cal R}$)}
\label{Sec:R}
\begin{figure} 
\begin{center}
\includegraphics[scale=0.3,angle=-90]{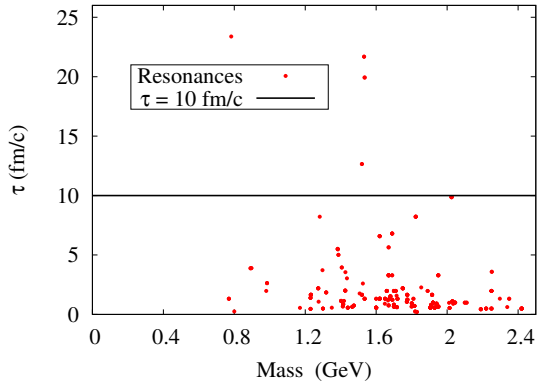}
\caption{(Color online) The values of mean life times (red points) for 
different hadron {\it resonances} up to $2.5$ GeV masses. 
The horizontal line indicates an approximated 
life time of the hadronic medium, produced in heavy ion experiments.
}
\label{particles_2}  
\end{center}
\end{figure}
Excluding the pseudo-scalar meson nonet and baryon octet,
the remaining part of hadronic zoo may be considered as ${\cal R}$ members.
The mean life times of ${\cal R}$~\cite{PDG} upto $M=2.5$ GeV are shown by red dots in 
Fig.~\ref{particles_2}, where a horizontal line indicates the life time of fireball,
which we have roughly considered as $10$ fm. The message of the plot is that
the ${\cal R}$ particles, having mean life times, less than the life time of medium, 
can only participate in the dissipation as they only can decay inside the medium.
Taking only those ${\cal R}$'s and using their mean life times as the relaxation 
times $\tau$ in Eqs.~(\ref{eta_G}) and (\ref{zeta_G}), we get the contribution of 
shear and bulk viscosities from ${\cal R}$ component.

%
%

\begin{figure}[ht]
\begin{center}
\includegraphics[scale=0.24,angle=-90]{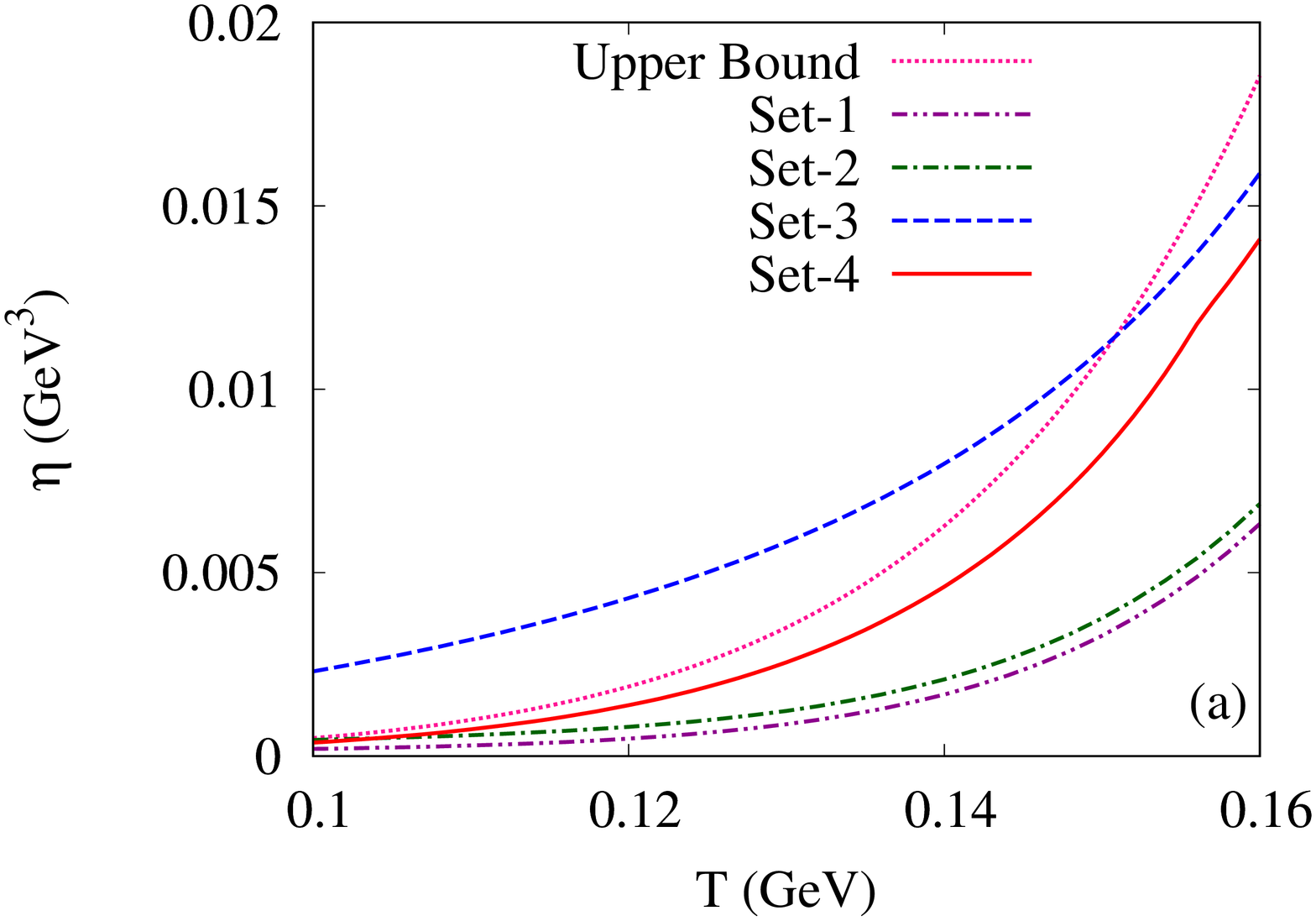}
\includegraphics[scale=0.24,angle=-90]{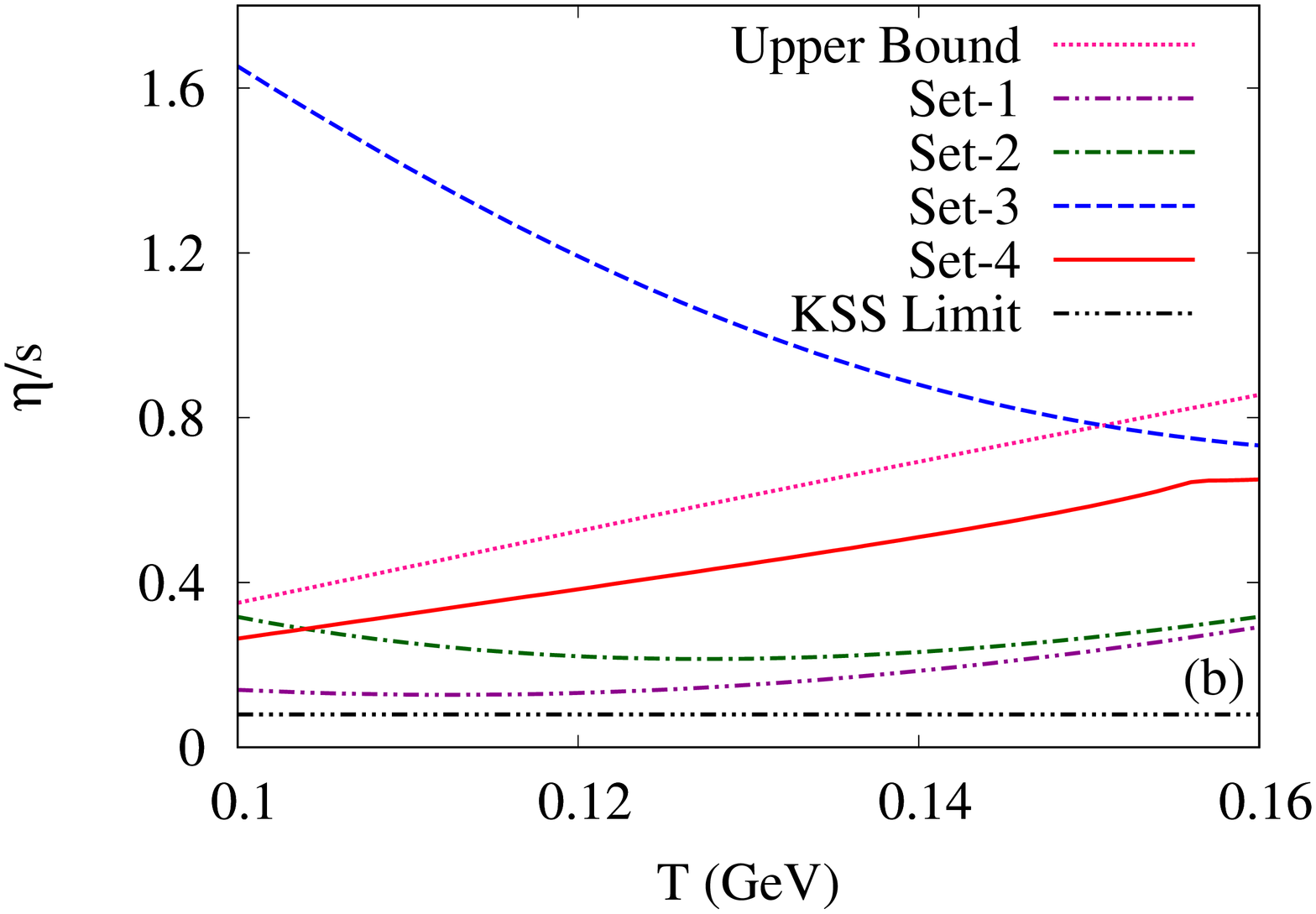}
\caption{(Color online) (a) Shear viscosity ($\eta$) (b) shear viscosity to entropy density ratio ($\eta/s$) 
	as a function of temperature for Set-1, 2, 3 and 4, as given in Table~(2). 
	An upper bound of both are shown by dotted and the dash-triple-dotted line indicates 
	the KSS limit for $\eta/s$.}
\label{fig:eta} 
\end{center}
\end{figure}

\begin{figure} 
\begin{center}
	\includegraphics[scale=0.24,angle=-90]{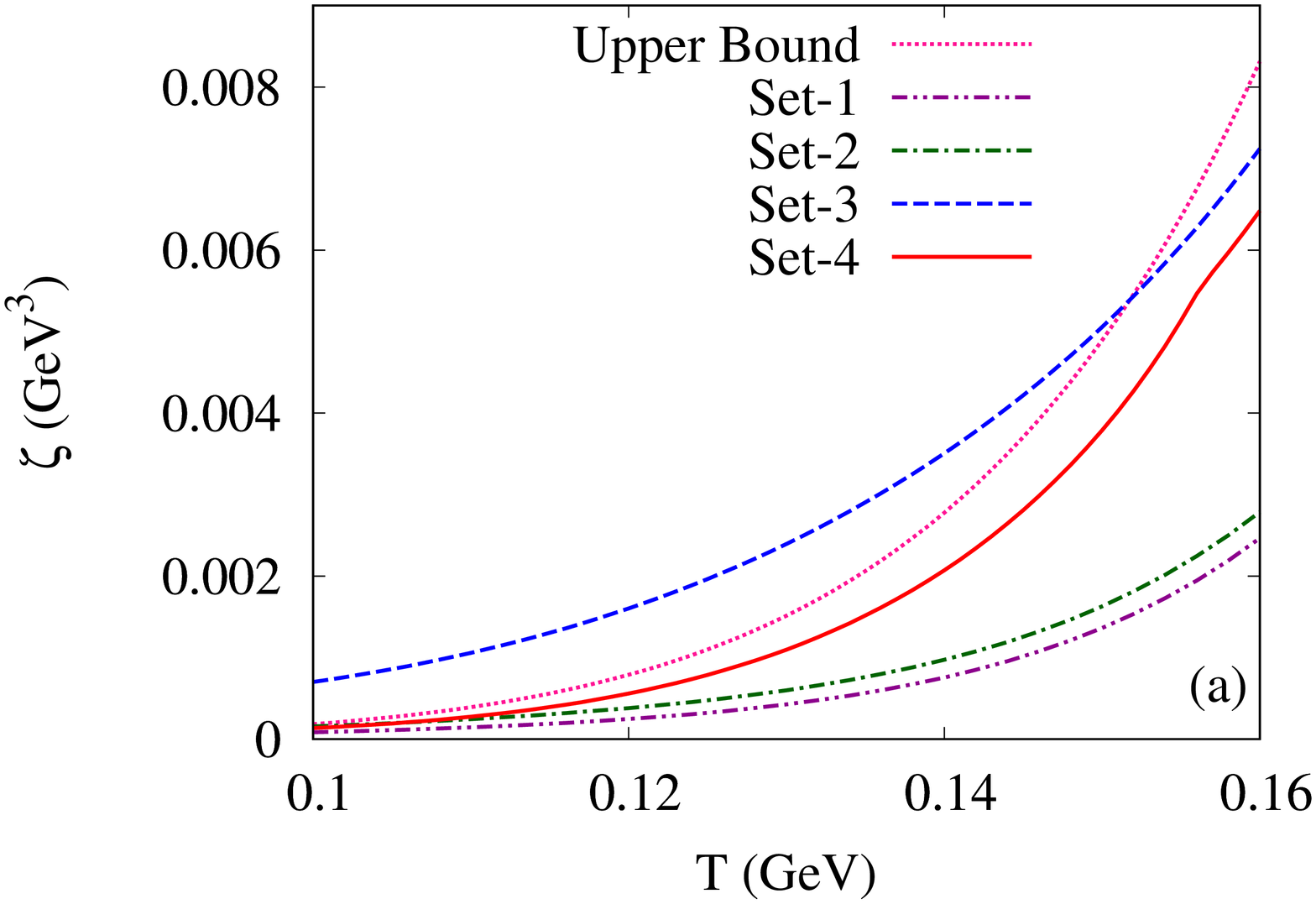}
	\includegraphics[scale=0.24,angle=-90]{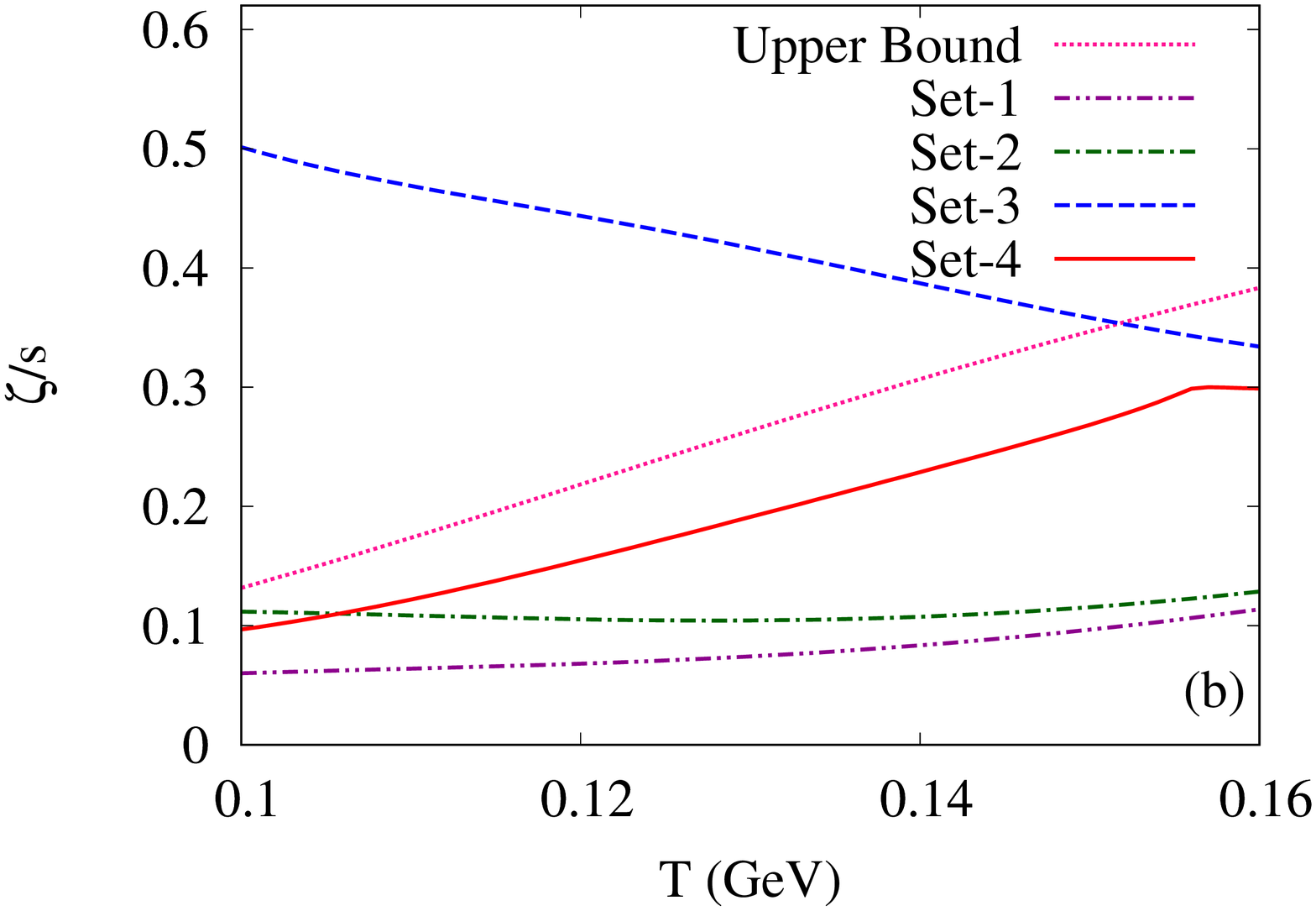}
	\caption{(Color online) The temperature dependence of (a) Bulk viscosity ($\zeta$) 
	(b) bulk viscosity to entropy density ratio ($\zeta/s$) for Set-1, 2, 3, 4 and 
	upper bound, as given in Table~(2).}
	\label{fig:zeta} 
\end{center}	
\end{figure}

Now, life time (maximum size) of fireball can be considered as upper
limit of relaxation times (relaxation lengths) of ${\cal NR}$ particles
and after adding this contribution with ${\cal R}$ part contribution, we
get an upper limit estimation of $\eta$ and $\zeta$, 
as shown by pink dotted line in Figs.~\ref{fig:eta}(a) and ~\ref{fig:zeta}(a). 
%

%

Normalizing the upper limit estimation of $\eta$ and $\zeta$ by the entropy density $s$, we get
a upper bound for $\eta/s$ and $\zeta/s$, as shown by pink dotted line in 
Figs.~\ref{fig:eta}(b) and ~\ref{fig:zeta}(b).
Here, we see that upper limit estimations for $\eta/s$ at $T=0.100$ GeV to $0.160$ GeV are
around $0.4$ to $0.85$, which is not very far from the KSS~\cite{KSS} or
quantum lower bound ($\frac{\eta}{s}=\frac{1}{4\pi}\approx 0.08$). It means that
a strongly coupled liquid behavior, which is experimentally expected in quark gluon plasma, 
is quite possible to see also in hadronic matter. 
The estimated values of $\eta/s$ at $T=0.100$ GeV to $0.160$ GeV in 
earlier Refs.~\cite{Nicola,Weise1,Itakura,Denicol,HM1,Bass,Juan}
are tabulated in Table~(1), which also contain the values of $\zeta/s$.
\begin{table} 
\begin{center}
\begin{tabular}{|l|c|c|}
\hline
  & $\eta/s$ at T=0.100 GeV  & $\eta/s$ at T=0.160 GeV \\
\hline
Upper bound~\cite{SG_HRG_PRC} & 0.4  & 0.85 \\
Fraile et al.~\cite{Nicola} & 0.9  & 0.3 \\
Lang et al.~\cite{Weise1} & 0.9  & 0.32 \\
Itakura et al.~\cite{Itakura} & 1.2  & 0.8 \\
Denicol et al.~\cite{Denicol} (HRG) & 1.4  & 0.12 \\
Kadam et al.~\cite{HM1} (HRG) & 1.1  & 0.2 \\
Demir et al.~\cite{Bass} (URQMD) & 1  & 1 \\
Rose et al.~\cite{Juan} (SMASH) & 1  & 1 \\
Set-1~\cite{Ghosh_pi,Ghosh_piN,SG_HRG_PRC} (HRG) & 0.13 & 0.28 \\
Set-2~\cite{SSS,SS_piN,SG_HRG_PRC} (HRG) & 0.32  & 0.31  \\
\hline
\hline
  & $\zeta/s$ at T=0.100 GeV  & $\zeta/s$ at T=0.160 GeV \\
\hline
Upper bound~\cite{SG_HRG_PRC} & 0.13  & 0.38 \\
Fraile et al.~\cite{Nicola} & 0.2  & 0.3 \\
Kadam et al.~\cite{HM1} (HRG) & 0.2  & 0.02 \\
Set-1~\cite{Ghosh_pi,Ghosh_piN,SG_HRG_PRC} (HRG) & 0.06  & 0.11 \\
Set-2~\cite{SSS,SS_piN,SG_HRG_PRC} (HRG) & 0.11  & 0.13 \\
\hline
\end{tabular}
\end{center}
\label{tab1}
\caption{Upper limit estimation and different model dependent
estimations of $\eta/s$ and $\zeta/s$ at $T=0.100$ GeV to $T=0.160$ GeV}
\end{table}

We notice that the values of $\eta/s$ and $\zeta/s$ near transition temperature or at $T=0.160$ GeV
remain approximately within the upper bound, estimated in HRG model for 10 fm system size but it
is not true for lower temperature, say $T=0.100$ GeV. Though upper bound of HRG model should not 
be considered as reference level of other model-calculations, but their low temperature results may be
modified if we properly impose the condition - the relaxation length of medium constituents should not 
be greater the system size. 
Refs.~\cite{Nicola,Weise1,Itakura} and Refs.~\cite{Bass,Juan} may not be relevant to compare quantitatively with 
values of upper bound because former references don't consider ${\cal R}$ members and latter references
have adopted models other than HRG. So, a direct quantitative comparison can be done with
HRG based works - Refs.~\cite{Denicol,HM1} where both have considered an approximate hard sphere
scattering strengths for all hadrons. This equal footing consideration for all hadrons are taken
for simplification but one should filter out the resonances, whose interaction time scales are 
larger than the life-time of the system ($\sim 10$ fm). Using this filtration for ${\cal R}$ members, as done 
in Ref.~\cite{SG_HRG_PRC} and using the relaxation time pion and nucleon 
from Refs.~\cite{Ghosh_pi,Ghosh_piN}, we have create a new set of estimation, which is
named as Set-1. In similar way Set-2 is created where the $\tau_\pi$ and $\tau_N$ results
are taken from Refs.~\cite{SSS,SS_piN}. Their results for $\eta$, $\eta/s$, $\zeta$, $\zeta/s$
are shown by dash-double-dotted and dash-dotted lines in 
Figs.~\ref{fig:eta}(a), (b), ~\ref{fig:zeta}(a), (b). None of the curves are crossing
their corresponding upper bound. It is expected because the relaxation lengths of neither ${\cal NR}$
members ($\pi$ and $N$) nor ${\cal R}$ members (due to proper filtration) exceed the value
of system size.
\subsection{Example of using upper momentum cut-off}
\label{Sec:Example}
%
%
Let us take an example whose values of $\eta/s$ and $\zeta/s$ 
cross our proposed upper bound. Here we will demonstrate how to consider
the dissipation of hadrons within the finite size hadronic matter
and how it can help to modify the values of transport coefficients within
our proposed upper bound. 
Using the experimental values of scattering lengths for
$\pi\pi$, $\pi N$, $NN$, $KN$ interactions from Refs.~\cite{exp2,exp3,exp1} 
and $\pi K$ interaction from Ref.~\cite{exp5,exp4}, their corresponding
relaxation lengths have been obtained in Ref.~\cite{SG_HRG_PRC}.
Using the full momentum distribution of $\pi$, $K$ and $N$ relaxation 
lengths in Eqs.~(\ref{eta_G}), (\ref{zeta_G}) and the filtered ${\cal R}$ contribution,
Ref.~\cite{SG_HRG_PRC} got one set of result, which is marked here as Set-3, given in Table.~(2).
The results of $\eta$, $\eta/s$, $\zeta$ and $\zeta/s$ are shown by blue dash line in 
Figs.~\ref{fig:eta}(a), (b), ~\ref{fig:zeta}(a) and (b), which are going beyond the 
our proposed upper bound in the low temperature domain.
The reason is hidden in momentum distribution of
$\pi$ and $K$~\cite{SG_HRG_PRC}, where their relaxation lengths at high momentum range exceed the value
10 fm, which is assumed as the size of our system.
So, we are counting some extra dissipation part, which is not at all contributing in 10 fm
system size.

\begin{table} 
\begin{center}
\begin{tabular}{|l|c|c|}
\hline
  & Transport coefficients  & Entropy density \\
\hline
{Upper bound} & ${\cal NR}$ ($\tau=10$ {fm}) + ${\cal R}$ ($\tau<10 {\rm fm}$)  & ${\cal NR}$ + ${\cal R}$ \\
\hline
{Set-1} &  ${\cal NR}$~$^{\rm Ref.}$~\cite{Ghosh_piN} + ${\cal R}$ ($\tau<10$ {fm}) & ''  \\
\hline
{Set-2} &  ${\cal NR}$~$^{\rm Ref.}$~\cite{SS_piN} + ${\cal R}$ ($\tau<10$ {fm}) & ''  \\
\hline
{Set-3}~$^{\rm Ref.}$~\cite{SG_HRG_PRC} & ${\cal NR}$ + ${\cal R}$ ($\tau<10$ {fm}) & ,, \\
\hline
{Set-4}~$^{\rm Ref.}$~\cite{SG_HRG_PRC} & ${\cal NR}$ ($\tau(\vk)<10$ {fm}) + ${\cal R}$ ($\tau<10$ {fm})   & ,, \\
\hline
\end{tabular}
\label{tab2}
\caption{Different set of inputs for transport coefficients ($\eta,~\zeta$) 
and entropy density (or other thermodynamical quantities like pressure, speed of sound)}
\end{center}
\end{table}

To resolve it, we have first track numerically the upper
momentum threshold or cut-off at different temperatures for $\pi$ and $K$, within which their
relaxation lengths don't exceed the fireball dimension (10 fm).
When we put those $T$ dependent
momentum thresholds as upper limit in integration of Eq.~(\ref{eta_G}), (\ref{zeta_G}) 
and use the modified results of $\pi$ and $K$, the total
values of $\eta$, $\eta/s$, $\zeta$ and $\zeta/s$ will politely 
go down below our proposed upper bound. 
This set is denotes as Set-4 in Table.~(2) and the results are shown 
by red solid lines in Figs.~\ref{fig:eta} and \ref{fig:zeta}.

Hence, our investigation states that the values of 
transport coefficients for hadronic matter will be within our proposed upper bound, 
when one will properly impose the finite size dissipation of ${\cal NR}$ and ${\cal R}$ components during
the calculation. 
\begin{figure}  
\begin{center}
	\includegraphics[scale=0.24,angle=-90]{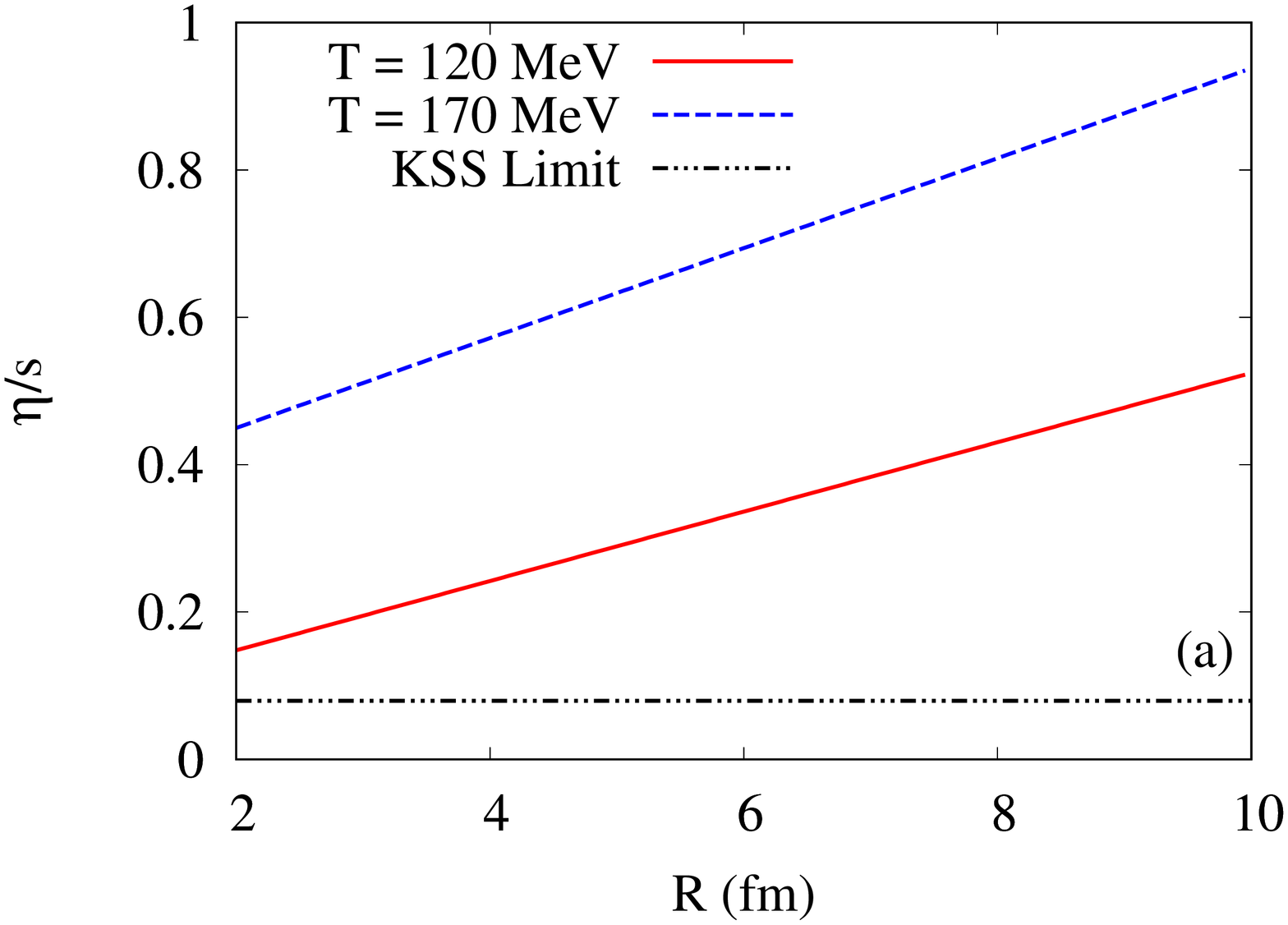}
	\includegraphics[scale=0.24,angle=-90]{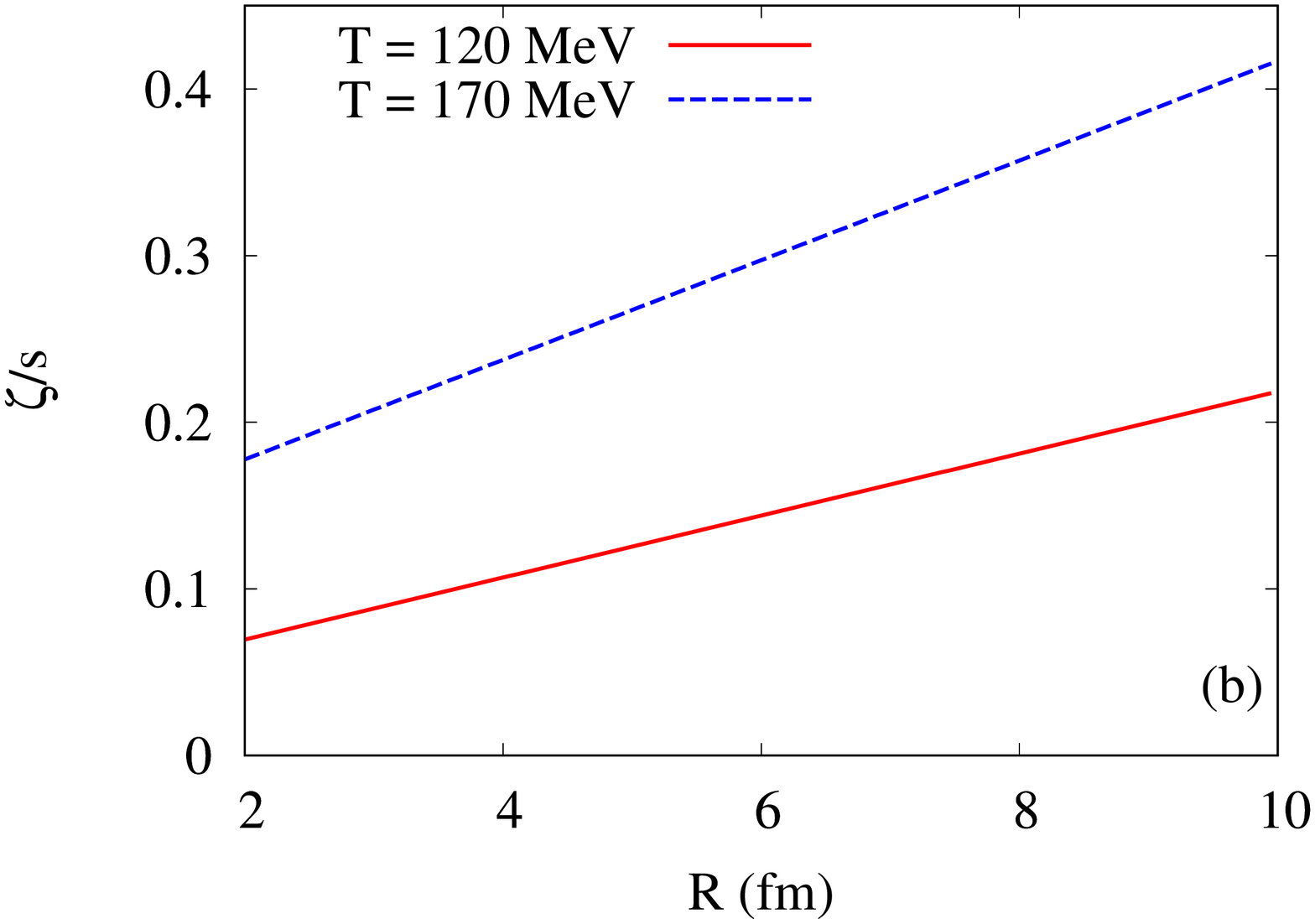}
	\caption{(Color online) Size (R) dependence of $\eta/s$ (a), $\zeta/s$ (b)
	at $T=0.120$ GeV and $T=0.160$ GeV.}
	\label{fig:eta_zeta_R} 
\end{center}	
\end{figure}

So far in Sec.~\ref{Sec:Relaxation}, we have deal with $R=10$ fm system size,
within which constituents are dissipating. Now if we decrease the $R$, then less
number of ${\cal R}$ members will take part in dissipation. So, considering those ${\cal R}$ members
and assuming $R$ as upper relaxation lengths of ${\cal NR}$ members ($\pi$, $N$ and $K$), we will
get $R$ dependent of upper bound for $\eta/s$, $\zeta/s$, as shown by red solid and blue dash lines in
Fig.~\ref{fig:eta_zeta_R} for $T=0.120$ GeV and $T=0.170$ GeV respectively. We notice that
reduction of system size can reduce the values of transport coefficients. In the context of fluid property, 
we can qualitatively conclude that $\eta/s$ of smaller medium can be more close to KSS bound.
It means that smaller size hadronic matter will be more close to a nearly perfect fluid behavior.

\section{Conclusions and Perspectives}
\label{Sec:Summary}
Present article has attempted to explore the microscopic calculations of
shear and bulk viscosities for hadronic matter, where ideal hadron resonance
gas model is used as a center tools. A special attention is drawn toward
the finite size effect of medium, for which one may get a phenomenological
upper bound of those transport coefficients. A broad numerical band,
within which earlier estimated values of transport coefficients are located,
can be shorten due to this finite size effect. Owing to the quantum effect,
finite size can introduce a lower momentum cut-off, from where medium
constituents will be distributed instead of starting from zero momentum.
We have found that both the thermodynamical quantities like entropy density $s$,
speed of sound $c_s$ and the transport coefficients like shear viscosity $\eta$,
bulk viscosity $\zeta$ can be changed when we reduce the system size lower than 6 fm.
%
%

The only dynamical input that goes in the calculation of $\eta$ and $\zeta$ is the relaxation times $\tau$
of different hadrons, which are classified into resonance $\mathcal{R}$ 
and non-resonance $\mathcal{NR}$ components based on the shorter and longer lifetimes with respect
to strong interaction time scale. The shorter mean life times of ${\cal R}$'s can be considered
as their relaxation times. Now, for a finite size (life time) of the hadronic medium, we have
to filter out those ${\cal R}$, whose relaxation length (time) cross the size (life time) of the medium.
On the other hand, relaxation time of ${\cal NR}$ members can be obtained from different
possible scattering processes, allowed by strong interaction Lagrangian densities. Again,
their relaxation length (time) should not exceed the system size (life time). Based on this
phenomenological restriction, one can get an upper bound of transport coefficients.
We have discussed about some sets of earlier results, where relaxation lengths of ${\cal NR}$ members
have not cross the system size, hence, their estimated values of transport coefficients remain
within our proposed upper bound. At the end, we have taken an example, whose relaxation lengths
of ${\cal NR}$ members at high momentum range cross the system size. Hence, its values of
transport coefficients cross our proposed upper bound. However, by introducing an upper
momentum cut-off of ${\cal NR}$ members, the values of transport coefficients are reduced
and gone below our proposed upper bound.


So, the current review article is desperately saying that 
any Hadronic model calculation of transport coefficients are different from estimation of those 
coefficients for 10 fm hadronic matter. Based on Refs.~\cite{SG_HRG_PRC,SG_HRGV_JPG}, 
a simplified lower momentum cut-off in thermodynamical phase-spase and upper momentum 
cut-off in relaxation length or filtering out higher decay lengths candidates are prescribed,
whose outcomes are quite inclined along the phenomenological expectation of transport coefficients
values. Therefore, this highly overlooked finite size effect framework should be improved towards
more realistic picture and the corresponding tools should also be applied in different alternative models.
It may be interesting to find the modified values of upper 
bound of $\eta/s$, $\zeta/s$ and its size dependence in the 
improved version of HRG models like excluded volume HRG (EV-HRG) 
models~\cite{Ev_HRG1,Ev_HRG2}, van der Waals (VDW) version of 
HRG models~\cite{VDW1,VDW2}.

\section*{Acknowledgements} Sabyasachi Ghosh thanks to Jayata Dey for useful help and 
gratefully acknowledges the contribution from his collaborators Sourav Sarkar, Bedangadas mohanty. 
Snigdha Ghosh acknowledges the Indian Institute of Technology Gandhinagar for the Post Doctoral Fellowship.

%
%
%

\end{document}